\documentclass[oldversion]{aa}
\usepackage[dvips]{graphicx}

\usepackage{rotating}
\usepackage{longtable}
\usepackage{amssymb,amsmath}
\newcommand{\be}{\begin{equation}}
\newcommand{\ee}{\end{equation}}
\usepackage{latexsym}
\newcommand{\simless}{\mathbin{\lower 3pt\hbox
      {$\rlap{\raise 5pt\hbox{$\char'074$}}\mathchar"7218$}}} %< or of order
\newcommand{\simgreat}{\mathbin{\lower 3pt\hbox
     {$\rlap{\raise 5pt\hbox{$\char'076$}}\mathchar"7218$}}} %> or of order

\newcommand{\sori}{$\sigma$~Ori}
\newcommand{\roph} {$\rho$~Oph}
\newcommand{\Msun}{M$_\odot$}

\newcommand{\Ha}{H$\alpha$}
\newcommand{\Pab}{Pa$\beta$}
\newcommand{\Pag}{Pa$\gamma$}
\newcommand{\Brg}{Br$\gamma$}

\newcommand{\Teff}{T$_{eff}$}
\newcommand{\Lstar}{L$_\ast$}
\newcommand{\Rstar}{R$_\ast$}
\newcommand{\Mstar}{M$_\ast$}
\newcommand{\Lacc}{L$_{acc}$}
\newcommand{\Macc}{$\dot M_{acc}$}

\begin{document}

\title{Accretion properties of  T Tauri stars in $\sigma$~Ori
\thanks{
Based on observations collected at the European Southern Observatory, Chile.  
Program 078.C-0382.}
}

\author{
T. Gatti\inst{1,2},
A. Natta\inst{1},
S. Randich \inst{1},
L. Testi\inst{1,3}
\and
G. Sacco \inst{4}
}

\institute{
    Osservatorio Astrofisico di Arcetri, INAF, Largo E.Fermi 5,
    I-50125 Firenze, Italy 
\and
Universit\`a  di Firenze, Dipartimento di Astronomia, Largo E.Fermi 5,
    I-50125 Firenze, Italy
\and
ESO, Karl-Schwarschild Strasse 2, D-85748 Garching bei M\"unchen, Germany
\and
Osservatorio Astronomico di Palermo, INAF, Piazza del Parlamento 1, 90134 Palermo, Italy
}

\offprints{natta@arcetri.astro.it}
\date{Received ...; accepted ...}

\authorrunning{Natta et al.}
\titlerunning{Accretion in $\sigma$~Ori}

\abstract
{Accretion disks around young stars evolve in time with time scales of few million years.  We present here a study of the accretion properties of a sample of
35 stars in the $\sim 3$ million year old star-forming region $\sigma$~Ori. Of these,
31 are objects with evidence of disks, based on their IR excess emission.
We use near-IR hydrogen recombination lines (Pa$\gamma$) to measure their
mass accretion rate. We find that the accretion rates are 
significantly lower
in $\sigma$~Ori than in younger regions, such as $\rho$~Oph, consistently
with viscous disk evolution. The He~I 1.083 $\mu$m line is detected (either in
absorption or in emission) in 72\% of the stars with disks, providing evidence
of accretion-powered activity also in very low accretors, where other accretion
indicators dissapear.}
%{bb}
%{cc}
%{dd}
%{ee}


\maketitle

\section {Introduction}

At the time of their birth, circumstellar disks are present around most (if not all) stars  of mass lower than few solar masses. 
The fraction of 
stars with disks decreases with time, from being close to 100\% in the youngest
regions to less than few percent after $\sim 5-7$ Myr (e.g., Hern\'andez et al.~\cite{Hea07} and references therein).
By the time a region is 10 Myr old, all the ``classical" (i.e., gas-rich, optically thick) disks have disappeared.
On similar timescales, all other indications of accretion-powered activity, such
as accretion and ejection of matter, disappear as well (e.g., 
Kenyon et al.~\cite{Kea05}; Mohanty et al.~\cite{Mea05}; Barrado y Navascu\'es 
and Mart\'in \cite{BNM03}).
These time scales are consistent to zero order with viscous disk evolution
(Hartmann et al.~\cite{Hea98}),
although many other processes may play a role in disk dissipation
(e.g., Hollenbach et al.~\cite{Hol00}).

Recently, the discussion on how disks evolve has gained new momentum
from the large surveys of star forming regions obtained with {\it Spitzer},
which provide measurements of the fraction of stars with disks based on well-defined, statistically significant samples. Particularly interesting is
the new class of 
``evolved"  disks, of which very few were known from ground-based
photometry (Skrutskie et al.~\cite{Skea90}). They are objects with IR excess emission weaker than 
typical classical T Tauri stars (CTTS); a fraction of these are
so-called  transitional disks,
i.e., objects with no near-IR excess, but excess emission at longer wavelengths.
The fraction of evolved disks is 
larger in older regions (Hern\'andez et al.~\cite{Hea07}). The weak emission
of evolved disks may be the result
of grain coagulation and settling (D'Alessio et al.~\cite{Dal06}, Dullemond and 
Dominik~\cite{Dul05}). Transitional objects are likely disks where small grains have been
cleared from the inner regions, and the discussion on their relevance for disk
evolution (do all disks disappear from inside out?)
is open.

Another indication of the presence  of disks is given
by  the accretion-powered phenomena that go under the label of accretion
activity (UV veiling, line emission and absorption from infalling and ouflowing gas, etc.). 
From them, it is possible to measure the mass accretion rate from the disk
onto the central object, a crucial quantity for constraining the physical properties of accretion disks.
Recently,
Sicilia-Aguilar 
et al.~(\cite{SAea06b}) measured the mass accretion rate in a number of stars of different age
from Taurus, \roph, Cha I, TW Hya and Tr~37
and found that \Macc\ decreases steadily with time, roughly as $t^{-1.5}$.
Such a trend is consistent with the  expectations of viscous disk evolution
(Hartmann et al.~\cite{Hea98}).

The sample used by
Sicilia-Aguilar et al.~(\cite{SAea06b}) is based on individual stellar ages
and
includes only few stars in  each region, with the
exception of Tr~37, which contributes more than half of the total.
The mass range of the stars with measured accretion rate in Tr~37 
(which has a distance of $\sim$ 900 pc) is shifted toward higher mass than the other, closer regions.
We think that it is important to
measure mass accretion rates in large numbers of stars in different
regions over a large range of average ages.
This is because
there is evidence that, in addition to the time dependence,
other parameters, not identified so far, are likley to affect disk evolution,
as shown  by the Muzerolle et al.~({\cite{Mea03}) and  Natta et al.~(\cite{Nea06})
studies of accretion in Taurus and Ophiuchus. These authors have
shown   that, firstly, 
the accretion rate depends on the mass of the central
object and that, at the same time,
there is a large range of disk and accretion properties even for
stars of similar mass and age in the same star forming region.
% which can have accretion rate and disk masses that vary by orders of magnitude.
Moreover, there is evidence that the disk lifetime is shorter for more
massive stars (Hillenbrand et al.~\cite{Hil98};
Carpenter et al.~\cite{Cea06}; Lada et al.~\cite{Lea06};
Hern\'andez et al.~\cite{Hea07}; Dahm and Hillenbrand \cite{DH07}). 
%The relation between this
%and the accretion dependence on mass is not known.

We are interested in  extending the studies of the accretion properties of young stars to older regions, to see how they change with time and, more specifically, if both the correlation between accretion rate and central mass and the
large spread of accretion rates remain unchanged in time.
In this paper, we present the first results of a study of the
accretion properties of CTTS in \sori.
The \sori\ cluster is 
located at a distance of $\sim 350$ pc and has an age of
$\sim 3$ Myr. It contains more than 300 stars, ranging in mass from the 
bright, massive multiple system \sori\ itself (spectral type O9.5) 
to brown dwarfs.
The \sori\ region has been extensively studied over the last 
few years in the optical, X-ray and infrared (e.g., Kenyon et al.~\cite{Kea05};
Zapatero-Osorio et al.~\cite{Zea02}; Franciosini et al.~\cite{Fea06};
Oliveira et al.~\cite{Oea04}, \cite{Oea06}; Caballero \cite{C07}; 
Caballero et al.~\cite{Cea07}). 
{\it Spitzer} has recently produced a complete census  
of the stellar population down to the brown dwarf regime (Hern\'andez et al.~\cite{Hea07}). These authors
find that the fraction of stars that retain classical (i.e., flared, gas-rich) disks (i.e., bona-fide Class II objects)
vary from about 10\% for stars with mass $>2$ \Msun\ to about 35\% for T Tauri stars (TTS) and brown dwarfs (BDs). 
A relatively large
fraction of TTS, about 15\%,  have  evolved disks; 
7 of these are
candidate transitional disks. The large and well characterized sample of
Class II and evolved disks in
\sori\ makes it particularly suited to study the properties of accretion and disks for an intermediate age stellar population.

We determine the accretion rate from the 
luminosity of the IR hydrogen recombination lines, \Pab\ and \Pag\ in 
particular. The reliability of this method has been discussed
by Muzerolle et al.~(\cite{Mea98}), Natta et al.~(\cite{Nea04}) and Calvet et
al.(\cite{Cea04}) and
applied by Natta et al.~(\cite{Nea06}) to a complete sample of IR-selected Class II stars in the young, embedded star forming region \roph.
Although for the optically visible objects in \sori\ other methods of measuring
the accretion rate could (and should) be used, we consider that 
the comparison between different regions is more significant if the same method
is used in all cases. 

We discuss the properties of our \sori\ sample  in \S 2, which also presents
details of the observations and data reduction. The results are presented in \S 3. The accretion rate measurements and their implications are discussed in \S 4. Summary and conclusions follow in \S 5.

\section{Sample, observations and data Reduction}

\subsection{The sample}

Our sample contains 35 TTS with evidence of accretion and/or disks.
We selected them from two different sources. The first
is Oliveira et al.~(\cite{Oea04}, \cite{Oea06}), who
detected a
K-L' color excess, very likely due to
 a circumstellar disk, in about  25 \sori\ TTS. The other source is the high spectral resolution spectral survey of
Sacco et al.~(\cite{Sea07}), who detected a total of 28
stars  with
broad H$\alpha$ lines (10\%FW$>$200 km/s), indicative of accretion. 
We imposed a limiting magnitude J$<$14.5, to achieve the required signal-to-noise.
Our sample includes 18 of the 23 Oliveira et al.
%~(\cite{Oea04}, \cite{Oea06})
sources above the J limit,  and 22/28 of the broad \Ha\ stars; 5 stars 
(number \#5, \#6, \#8, \#14 and \#22 in Table~\ref{table_1}) are in common
(they have both K-L' color excess and strong
 H$\alpha$), for a total of 35 objects.
Sacco et al.~(\cite{Sea07})
confirm the membership of the
22 stars in their sample; 
none is found to be binary.

After completion of our observations, 
Hern\'andez et al.~(\cite{Hea07}), published  the results of their
{\it Spitzer} survey of the \sori\ region.
Of our 35 objects,  32 lie in the surveyed field. 
Based on the shape of their spectral energy distribution (SED) in the near and mid-IR, 26 of them are classified 
as Class II, or classical TTS (CTTS), i.e., objects with gas-rich, flared disks
extending close to the star;
4 are Class III (i.e., TTS with no evidence of disk)
and 2 (\#28 and \#29) are evolved disks, i.e.,
objects with IR excess weaker than typical CTTS.
In the following, we will analyze them together with the Class II objects
and use the definition of Class II for the whole disk sample
(but see \S 4.4).
The SED classification is given in Table \ref{table_1}.
%the 28, a transitional disk candidate classified as Evolved Disk, the 29, a Class II classified as Evolved Disk and the 30, a Class II classified as Class I in ***. 
Three stars, \#1, \#13, and \#33, are not in the fields observed by Hern\'andez et al.~(\cite{Hea07}). We classify them as Class II based on the presence of a K-L' 
excess (Oliveira et al.~\cite{Oea04}; \cite{Oea06}); in one case (\#1) the excess is at the 2.5$\sigma$ level, and the classification uncertain.

Within our J-magnitude limit (J$\leq$14.5),
our sample of disk objects contains about 30\% of the \sori\ members with evidence of disks from the {\it Spitzer} survey. 
Their distribution in color and magnitude is very similar to that
of the parental population (see Fig.~\ref{fig_1}), so that
we consider our total sample of 31 Class II
objects to be representative of the whole disk population.
%Results derived from our sample should be valid in general.
% to be used in the study of accretion properties of the whole star forming region, from a statistical point of view.

\begin{table*}
\caption {Stellar Parameters and Observed Properties}
%\begin{tiny}
\begin{center}
\begin{tabular}{ccccccccccccccc}
\hline 
\multicolumn{1}{c}{Obj} &
\multicolumn{1}{c}{Spitz.} &
\multicolumn{1}{c}{RA} &
\multicolumn{1}{c}{DEC} &
\multicolumn{1}{c}{J} &
\multicolumn{1}{c}{ST} &
\multicolumn{1}{c}{Cl.} &
\multicolumn{1}{c}{L$_{*}$} &
\multicolumn{1}{c}{M$_{*}$} &
\multicolumn{1}{c}{T$_{eff}$} &
\multicolumn{1}{c}{Pa$\beta$} &
\multicolumn{1}{c}{Pa$\gamma$} &
\multicolumn{1}{c}{HeI} &
\multicolumn{1}{c}{L$_{acc}$} &
\multicolumn{1}{c}{$\dot{M_{acc}}$}\\

\multicolumn{1}{c}{\#} &
\multicolumn{1}{c}{\#} &
\multicolumn{1}{c}{(2000)} &
\multicolumn{1}{c}{(2000)} &
\multicolumn{1}{c}{(mag)} &
\multicolumn{1}{c}{} &
\multicolumn{1}{c}{} &
\multicolumn{1}{c}{(L$_{\odot}$)} &
\multicolumn{1}{c}{(M$_{\odot}$)} &
\multicolumn{1}{c}{(K)} &
\multicolumn{1}{c}{($\AA$)} &
\multicolumn{1}{c}{($\AA$)} &
\multicolumn{1}{c}{($\AA$)} &
\multicolumn{1}{c}{(L$_{\odot}$)} &
\multicolumn{1}{c}{($M_{\odot}/yr$)}\\

\hline \hline

1  & -- & 5:37:56.1& -2:09:26.7 & 13.90& M4.5  & II  &  0.04    & 0.16  &  3075    & $<$0.5      & $<$0.6       & $-$       &  $<$-3.5  &  $<$-10.4    \\  
2  & 283 &5:37:58.4& -2:41:26.2& 13.28&M5.0  & III &  0.07    & 0.12  &  3000    & $<$0.5      & $<$0.4       & $-$       &  $<$-3.4  &  $<$-10.0    \\
3  & 451&5:38:18.9& -2:51:38.8 & 12.83& M3.0  & II &  0.12    & 0.25  &  3350    & $<$0.5      & 2.0$\pm$0.4  & 1.7$\pm$0.3  &    -2.2        &     -9.1    \\
4  & 462& 5:38:20.5& -2:34:09.0 & 12.65& M4.0  & II &  0.13    & 0.16  &  3150    & $<$0.4      & $<$0.3       & $-$       &  $<$-3.2      &     $<$-9.8    \\
5  & 518& 5:38:27.3& -2:45:09.7 & 11.95& K8.0  & II &  0.42    & 0.5   &  3900    & 9.4$\pm$0.5 & 6.7$\pm$0.6  & 7.3$\pm$0.5  &     -1.0  &      -8.1    \\
6  & 520 & 5:38:27.5& -2:35:04.2 &12.83& M3.5  & II &  0.12    & 0.2   &  3250    & $<$0.6      & 0.8$^c$  & $-$  &    -2.8      &      -9.6    \\ 
7  & 562& 5:38:31.4& -2:36:33.8 & 12.17& M3.0  & II &  0.19    & 0.2   &  3350    & 4.9$\pm$0.4 & 3.6$\pm$0.3  & 4.7$\pm$0.3  &      -1.5       &      -8.2    \\
8  & 563 & 5:38:31.6& -2:35:14.9 & 11.52& M2.0  & II &  0.23    & 0.3   &  3500    & $<$0.5     &  $<$0.5       & 1.8$\pm$0.5  &   $<$-2.3       &   $<$-9.2    \\ 
9  & 592& 5:38:34.3& -2:35:00.1 & 11.22& K0.0  & III &  0.16    & 0.8   &  5250    & $<$0.4     &  $<$0.4       & $-$      &   $<$-2.3     &     $<$-10.0    \\
10 & 598 & 5:38:34.6& -2:41:08.8& 13.10& M4.5  & II &  0.11    & 0.14  &  3075    & $<$0.6     &  $<$0.7       & -0.8$^c$      &   $<$-3.0      &    $<$-9.6    \\
11 & 611& 5:38:35.5& -2:31:51.7 & 11.30& K7.0  & III &  0.57    & 0.5   &  4000    & $<$0.7     &  $<$0.7       & $-$       &   $<$-2.0    &      $<$-9.0    \\
12 & 646& 5:38:39.0& -2:45:32.2 & 12.91& M3.5  & II &  0.10    & 0.2   &  3225    & $<$0.4     & 0.8$\pm$0.3   & -1.8$\pm$0.4 &     -2.8      &       -9.6    \\ 
13 & --  & 5:38:39.8& -2:56:46.2& 11.43& M0.5  & II &  0.45    & 0.4   &  3725    & $<$0.7     & $<$0.4        & 2.8$\pm$0.3  &   $<$-2.4       &   $<$-9.3    \\ 
14 & 662& 5:38:40.3& -2:30:18.5 & 11.51& M0.5  & II &  0.44    & 0.4   &  3725    & $<$0.7     & 0.7$^c$   & -0.7$^c$      &    -2.1      &       -9.0    \\
15 & 669 & 5:38:41.3& -2:37:22.6& 11.46& M0.0  & III &  0.44    & 0.45  &  3800    & $<$0.7     & $<$0.4       & 0.7$^c$   &  $<$-2.4       &   $<$-9.4    \\
16 & 682 & 5:38:42.3& -2:37:14.7& 11.77& M1.5  & II &  0.35    & 0.3   &  3575    & $<$0.4     & $<$0.5       & -1.2 $\pm$0.4       &   $<$-2.5      &   $<$-9.3    \\
17 & 694 & 5:38:43.9& -2:37:06.8& 12.84& M4.5  & II &  0.1     & 0.14  &  3075    & $<$0.4     & $<$0.4       & 1.0$\pm$0.3  &    $<$-3.2      &    $<$-9.8    \\
18 & 697 & 5:38:44.2& -2:40:19.7& 11.36& K7.5  & II &  0.63    & 0.45  &  3950    & $<$0.5     & $<$0.3       & 2.1$\pm$0.3  &    $<$-2.5       &   $<$-9.5    \\
19 & 710& 5:38:45.4& -2:41:59.4& 11.99& M2.5  & II &  0.25    & 0.25  &  3425    & $<$0.6     & 1.3$\pm$0.3  & 0.3$^d$  &    -2.0       &      -8.8    \\  
20 & 723 & 5:38:47.2& -2:34:36.8& 12.56& M3.5  & II &  0.09    & 0.2   &  3250    & 3.4$\pm$0.5 & 2.1$\pm$0.4  & $-$      &      -2.0      &       -8.9    \\ 
21 & 726& 5:38:47.5& -2:35:25.2 & 11.74& M1.5  & II &  0.42    & 0.3   &  3575    & 1.1$\pm$0.3 & 1.4$\pm$0.3  & $0.0^e$      &   -1.8   &         -8.6    \\ 
22 & 733 & 5:38:47.9& -2:37:19.2& 12.02& M2.5  & II &  0.36    & 0.25  &  3425    & $<$0.4     & 1.3$\pm$0.3  & 0.6$^c$      &      -2.0   &         -8.7    \\
23 & 736 & 5:38:48.0& -2:27:14.2& 10.16& K8.0  & II &  0.83    & 0.45  &  3950    & $<$0.6     & $<$0.5       & -1.5$\pm$0.4 &    $<$-1.6     &    $<$-8.5    \\
24 & 750 &5:38:49.3& -2:23:57.6& 14.36& M4.5  & II &  0.03    & 0.16  &  3075    & $<$1.2     & $<$1.2       & $-$       &    $<$-3.4      &   $<$-10.3    \\
25 & 762 &5:38:50.6& -2:42:42.9& 13.84& M5.0  & II &  0.04    & 0.12  &  3000    & $<$0.5     & $<$0.7       & 2.9$\pm$0.4  &    $<$-3.4      &   $<$-10.1    \\
26 & 774 & 5:38:52.0& -2:46:43.7 & 11.52& K8.0  & II &  0.44    & 0.5   &  3900    & $<$0.5     & 1.6$\pm$0.5  & 5.2$\pm$0.5  &    -1.6       &      -8.7    \\
27 & 871 & 5:39:04.6& -2:41:49.4& 13.96& M2    & II &  0.1    & 0.3    &  3500     & $<$1.2     & $<$1.0       & 8.5$\pm$1.0  &    $<$-3.2      &   $<$-10.3  \\  
28 & 897 & 5:39:07.6& -2:32:39.1& 11.30& K8.0  & II$^a$& 0.6& 0.5& 3900 & $<$0.7     & $<$0.6       & 1.2$^c$  &    $<$-2.1 &     $<$-9.1    \\
29 & 908 & 5:39:08.8& -2:31:11.5&13.07& M4.5  & II$^b$&  0.07& 0.16& 3075 & 1.3$\pm$0.4 & 0.8$\pm$0.3 & -1.3$\pm$0.3 &      -2.9      &       -9.6    \\   
30 & 927 & 5:39:11.5& -2:31:06.6& 11.99&  M0.5  & II&  0.24 &0.5  &3725   & $<$0.8     & $<$0.7       & $-$      &     $<$-2.4      & $<$-9.5    \\
31 & 984 & 5:39:18.8& -2:30:53.1& 11.40& K9.0  & II &  0.52    & 0.45  &  3850    & $<$0.6     & $<$0.4       & -1.1$\pm$0.3 &    $<$-2.4       &   $<$-9.4    \\
32 &1152 & 5:39:39.4& -2:17:04.5& 11.67& M1.5  & II &  0.29    & 0.35  &  3575    & $<$0.6     & $<$0.5       & -2.4$\pm$0.5 &    $<$-2.4       &   $<$-9.3    \\
33 & --   & 5:39:41.0& -2:16:24.4& 12.91& M4.5  & II &  0.10    & 0.14  &  3075    & $<$0.4     & $<$0.4       & $-$      &     $<$-3.2      &   $<$-9.8    \\
34 &1230 & 5:39:49.4& -2:23:45.9& 13.44& M4.0  & II &  0.07    & 0.18  &  3150    & $<$0.8     & $<$0.8       &0.8 $^c$      &    $<$-3.1      &   $<$-9.9    \\
35 &1248 &5:39:51.7& -2:22:47.2& 12.60& M5.0  & II &  0.13    & 0.12  &  3000    & $<$0.5     & $<$0.6       & $-$      &    $<$-2.8      &   $<$-9.3    \\

\hline             
%\hline
%\end{tiny}
\label{table_1} 
\end{tabular}
%\linespread{1.0}
\end{center}
Notes: $a$, $b$: Transition disk and evolved disk, respectively, in Hern\'andez et al.~(\cite{Hea07}). $c$: tentative (2$\sigma$) detection. $d$: Type IVr P-Cygni profile: EW(em)=1.2$\pm$0.3, EW(abs)=-0.9$\pm$0.3.  $e$: tentative type IVr P-Cygni profile (0.6, -0.6).
\end{table*}
%\end{sidewaystable}

%\begin{sidewaystable}

\begin{figure}
   \centering
   \resizebox{\hsize}{!}{\includegraphics[]{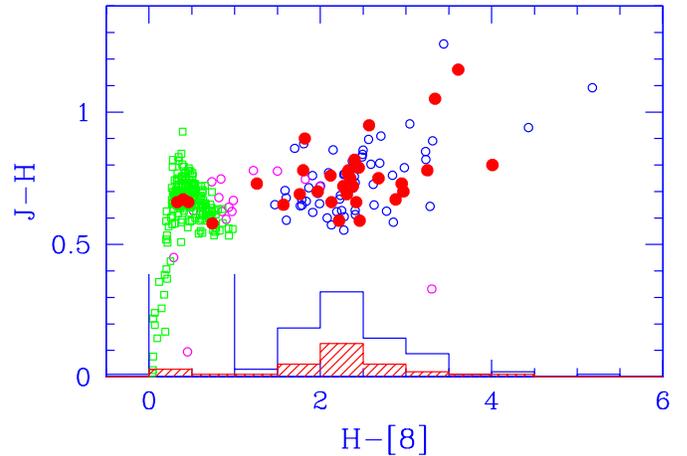}}
	\caption{ \label{fig_1} [J-H] versus H-[8] color diagram for 
\sori\ cluster members with J$<$ 14.4. Empty circles and squares are Class II/transitional disks  
and Class III objects respectively. Red filled circles show the location of our \sori\ sample with Spitzer observations (32 objects in total).
The histograms at the bottom show the distribution in H-[8] of all cluster members
(empty) and of our sample (dashed), respectively. All Class II disks have
H-[8] $> 1.5$.
}
\label {fig_colors}
\end{figure}

\begin{figure}
   \centering
   \resizebox{\hsize}{!}{\includegraphics[]{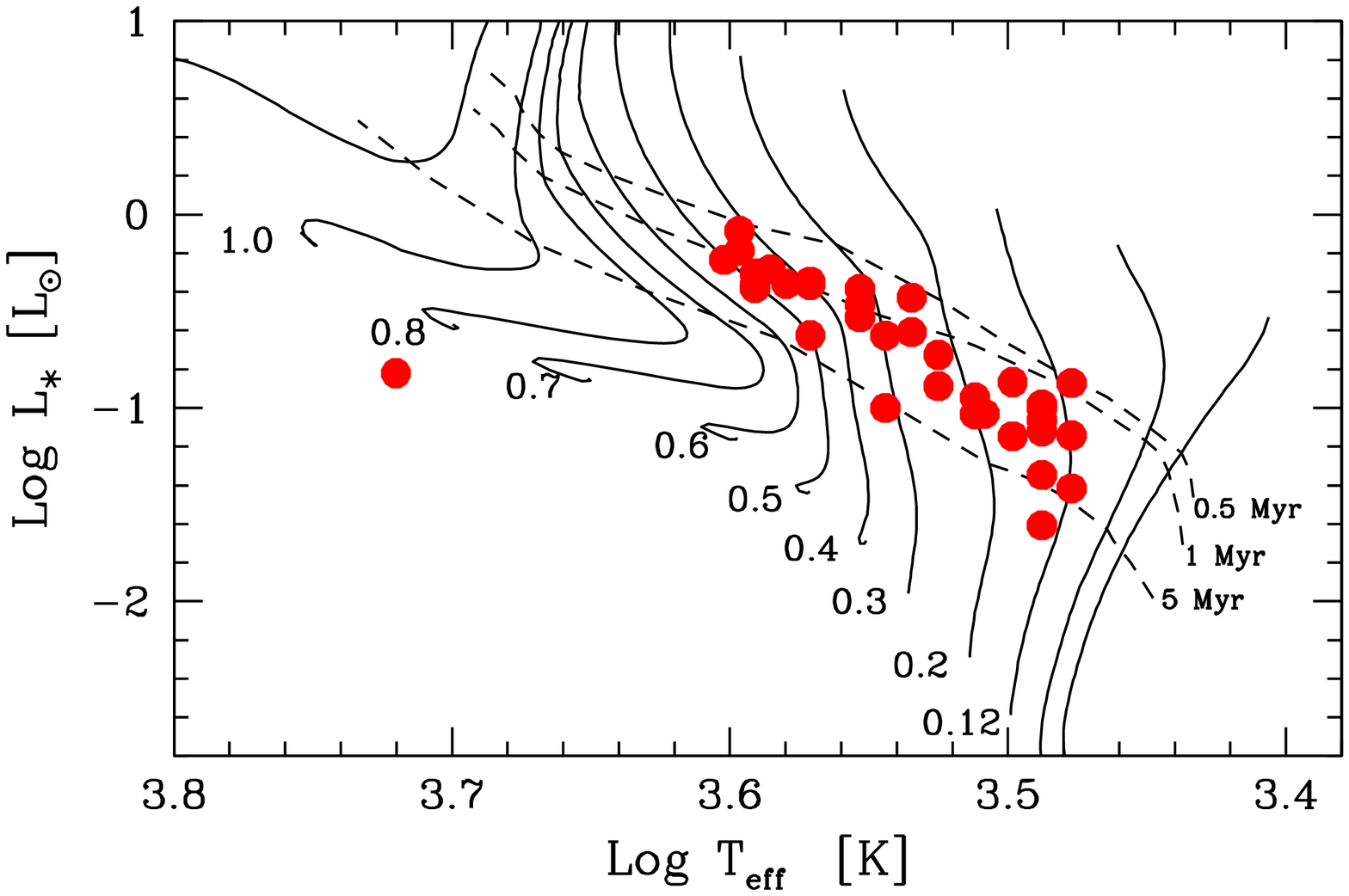}}
	\caption{ \label{fig_2} Location of the \sori\ targets
on the HR diagram. Evolutionary tracks from D'Antona and Mazzitelli (\cite{DM97}), labelled with the stellar mass. Isochrones for 0.5, 1 and 5 Myr are shown
by the dashed lines. }
\label {fig_hr}
\end{figure}

%%\end{sidewaystable}

\subsection{Stellar parameters}

Spectroscopically determined spectral types exist for a small fraction of
\sori\ objects only (e.g., Franciosini et al.~\cite{Fea06} and references therein). To ensure homogeneity, we  determine the effective temperature of each object from its visual photometry, adopting the 
relation between color indexes,  \Teff\ and spectral types
of Bessell~(\cite{Bes91}) for M dwarfs stars, and from Bessell~(\cite{Bes79}) and Bessell and Brett~(\cite{Bes_bret88}) for stars of earlier spectral type. We assume negligible extinction in all bands (Brown et al.~\cite{Bro94}; B\'ejar et al.~\cite{Bej99}), and compute
\Teff\ independently from the three
color indexes (V-R), (V-I), (R-I). The optical photometry is
from Sherry et al.~(\cite{She04}), Zapatero-Osorio et al.~(\cite{Zea02}), Kenyon et al.~(\cite{Kea05}), B\'ejar et al.~(\cite{Bej01}) and Wolk (\cite{Wolk96}). 
The differences between the different determinations
of \Teff\ are generally
less than  $\pm$ 150 K, and often much smaller. 
The adopted values of \Teff\ and spectral types
are given in Table \ref{table_1}.
The comparison with the spectroscopic determinations of  spectral types
of Zapatero-Osorio et al.~(\cite{Zea02}) 
for the 8 objects in common shows differences of 1/2 to 1 subclass.

Stellar luminosities  have been computed from the I magnitude, using the
bolometric correction appropriate for the spectral type (Kenyon and Hartmann ~\cite{KH95}),
no extinction and a distance D=350 pc. 
The location  of the stars on the HR diagram is shown in Fig.~\ref{fig_hr},
together with the evolutionary tracks of D'Antona and Mazzitelli (DM97;
\cite{DM97}).
The distribution of the stars is
consistent with a  typical age of $\sim 3$  Myr (Oliveira et al.~\cite{Oea04}),
with the exception of  \#9,
a Class III object with color indexes consistent with an early K 
spectral type but  very low luminosity.
%In general, 
%the uncertainties on \Lstar\ are dominated by the uncertainty on the \sori\ distance.
%If the region is at a distance of 440 pc, as adopted, e.g.,
%by Hern\'andez et al.~(\cite{Hea07}), the luminosities will be higher for all objects by 
%0.2 dex, and the age of the region closer to 1 Myr.

The stellar masses range from
$\sim 0.5$ to $\sim 0.1$ \Msun. 
Note that the exact values of \Mstar\ depend on the adopted
evolutionary tracks, especially around 0.5-1 \Msun. We use  DM97
for homogeneity with the analysis of Natta et al.~(\cite{Nea06})
of the mass accretion rates in \roph.
The lower mass limit is determined by our selection criterion that J$<$14.5 mag; the lack of more massive stars reflects the strong
decrease of the fraction of Class II objects for increasing J magnitude 
observed  by Hern\'andez et al.~(\cite{Hea07}). 
This trend,  seen in other star forming regions 
(e.g., Hillenbrand et al.~\cite{Hil98};
Carpenter et al.~\cite{Cea06}; Lada et al.~\cite{Lea06};
Hern\'andez et al.~\cite{Hea07}; Dahm and Hillenbrand \cite{DH07}), is
particularly strong in \sori, where only 4 (7\%) stars brighter than J=11 mag are classified as Class II. One of them is included in our sample (\#23), 
and has an estimated spectral type K8 and mass $\sim 0.5$ \Msun. 
The fact that no higher-mass member 
has an associated Class II disk is confirmed
by Caballero (\cite{C07}): of the
18 very bright stars (spectral type G0 and earlier) 
in the \sori\ cluster studied by him
and included in the list of cluster members by
 Hern\'andez et al.~(\cite{Hea07}), 3 have 
transitional or debris disks, all the others are Class III.

%Finally we have put our objects in the HR diagram and we have obtained masses for all our stars, except for 27, from D'antona Mazzitelli evolutionary tracks (see Fig.\ref{fig4}). The result is a sample of stars covering omogenuosly the range 0.1 M$_{\odot}$ 0.5 M$_{\odot}$, where all the members, except the 24, range in age between 0.5 Myr and 5 Myr, where most objects are along the 3 Myr isochron, as the age commonly indicated for this star forming region (Ref).

%%\end{sidewaystable}

\subsection{Observations}

Near-infrared spectra in the J band were obtained for all our targets using SOFI near-infrared camera and spectrograph at the ESO-NTT telescope. The observations of \sori\ were carried out in Visitor Mode on December 1-4, 2006. The 0.6 arcsec slit and the Blue low resolution grism have been used, resulting in a spectral resolution $\lambda/\Delta\lambda \sim$ 1000 and a spectral coverage from $\sim$0.95 to $\sim$1.64$\mu$m. Integration time was 40 minutes for all the targets;
standard calibrations (flats and lamps) and telluric standards spectra were obtained for each observation. Wavelength calibration was performed using the lamp observations.

The mean seeing during the three nights was 0.7 arcsec, with many hours below 0.5 arcsec.
% so we have employed the 0.6 arcsec slit without important flux losses. 
All target objects and telluric standards have been observed at airmass values z$<$2.

With the goal of observing as many objects as possible, the targets have been observed in pairs, choosing stars with angular separations which allow a 20 arcsec nodding and a 6 arcsec jitter with the 4.71 arcmin slit, and orienting the slit at
 appropriate position angles. In total, we have observed in this way
24 targets, while 11 have been observed individually. 
%We have been able to make 12 "double pointments" (24 targets), we have observed 12 single stars, and all the telluric standards many times during every night. 
%These continuum fluxes have
%uncertainties of 20\% at least, deriving from the fact that, in order to maximize the number of objects we observed, we have not aligned the slit along the 
%parallactic axys. 
%During the acquisitions, the counting level $N$ achieved has been monitored continuosly, and $N$ values has been always below 10000, the limit at which the
%SOFI response becomes non-linear.

The spectra have been reduced using standard procedures in IRAF.
We are interested in this paper in the three lines indicative of accretion powered activity which fall in the J band, namely
the  HeI line at 1.083 $\mu$m and  the two hydrogen recombination lines
\Pag\ at 1.094 $\mu$m and \Pab\ at 1.282 $\mu$m; the portions of the spectra which contain these lines are shown in 
Fig.~\ref{fig_pag} and \ref{fig_pab}. 
Their  equivalent widths are given in Table \ref{table_1} (positive values for emission lines). Upper limits to
\Pab\ and \Pag\ assume that the lines are in emission; 
they have been estimated from the peak-to-peak noise in the spectral region 
of the lines, under the assumption that the lines are not resolved.
Since the He line can be either in emission or in absorption or both, we do not explicitely give upper limits in Table \ref{table_1},  
they are similar to those of the adjacent \Pag\ line.
%was measured
% by fitting Gaussian profiles to the spectra using standard IRAF routines
%upper limits are the bigger of the peak-to-peak
%noise of the continuum adjacent to the
%line and the equivalent width of the noise features, in a manner similar to
%what had been done for \roph\ by Natta et al.~(\cite{Nea06}). 
%The  results are given in Table ??.

\section {Results}

The presence of the hydrogen  lines in emission and of the He~I
line, often with   P-Cygni profiles clearly seen in high resolution spectra, (Edwards et al.~\cite{Eea06}) 
is  evidence of accretion related activity. In fact,
inspection of Table \ref{table_1}
 shows that none of the three 
lines is detected in Class III stars, with the possible exception of 
\#15, where we have a tentative detection of He~I in emission.
Of the 31 Class II, we detect at least one of the three lines in 26 cases.
The  hydrogen lines are detected in emission in 12 Class II stars, but only in 5 we do detect both of them; in 7 cases, only \Pag\ is clearly seen.
The hydrogen line  detection is confirmed in 10/12 cases by the presence of the He~I
line, either in absorption or in emission.

\begin{figure*}
   \centering
   \resizebox{\hsize}{!}{\includegraphics[]{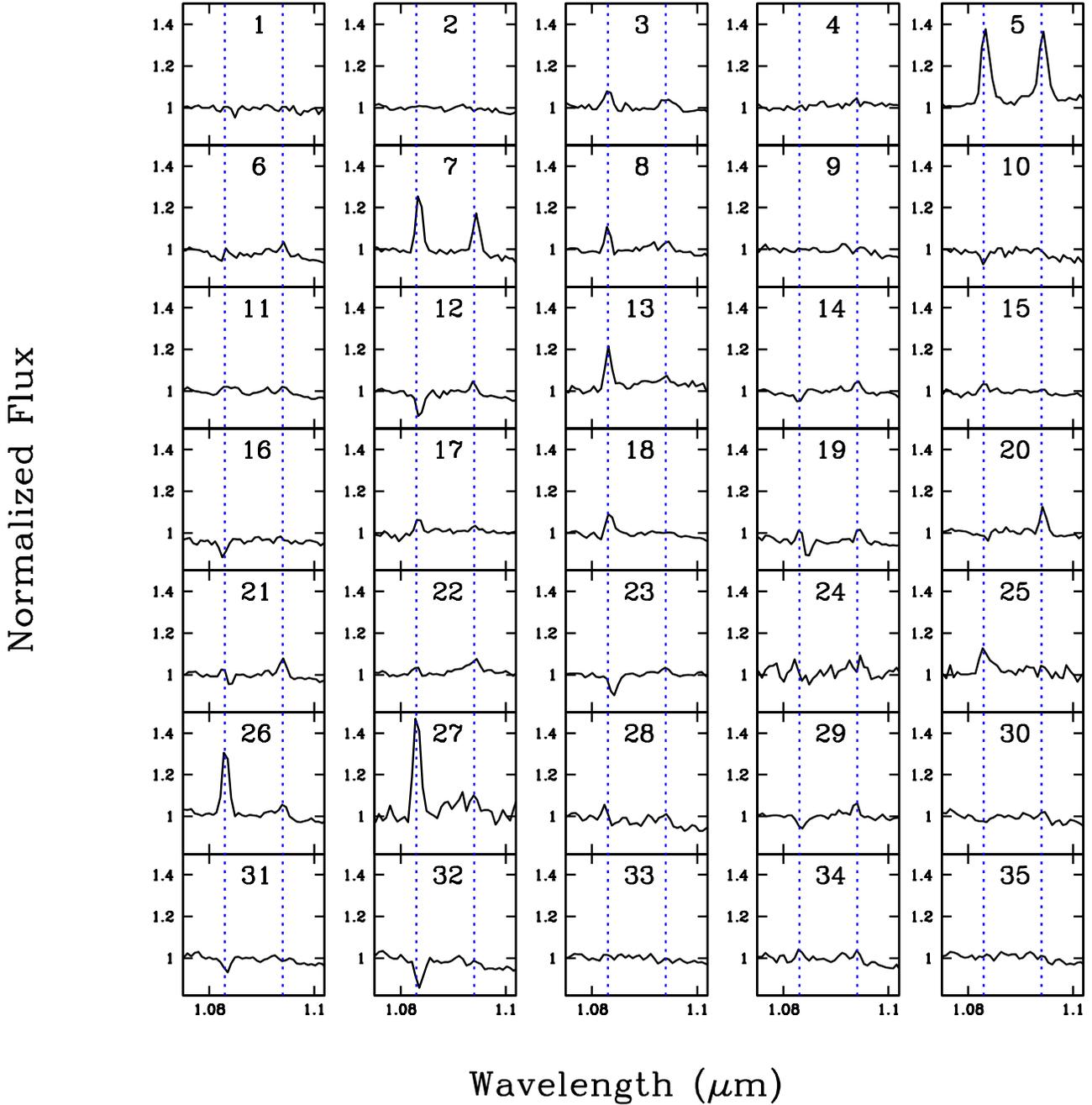}}
	\caption{ Normalized observed spectra in the HeI and \Pag\ region.
The dotted lines show the wavelengths of the two lines.}
\label {fig_pag}
\end{figure*}

\begin{figure*}
   \centering
   \resizebox{\hsize}{!}{\includegraphics[]{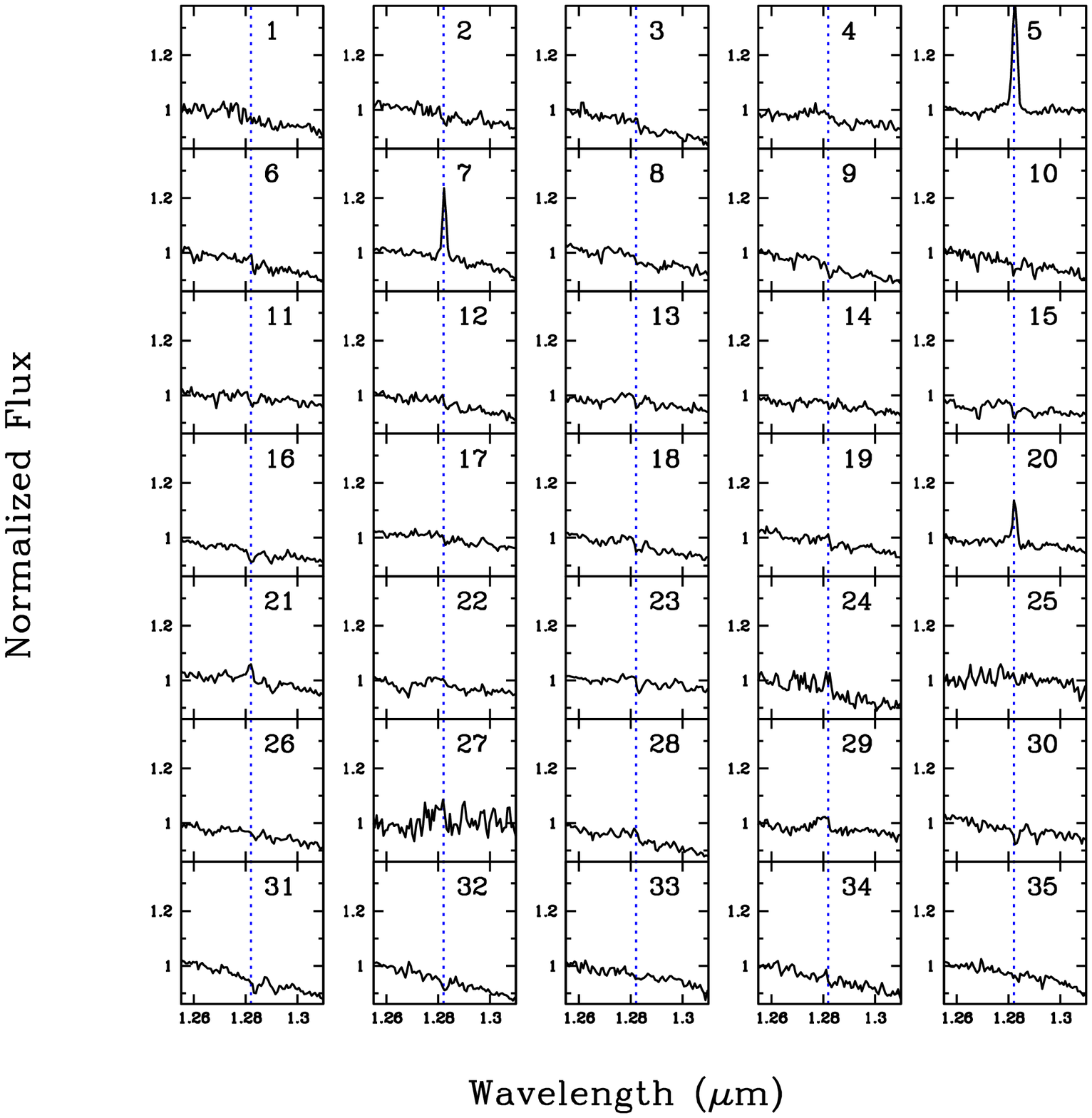}}
	\caption{Normalized observed spectra in the \Pab\ region. The wavelengtth of \Pab\ is shown by the dotted line. }
\label {fig_pab}
\end{figure*}

Our spectral resolution ($\sim$ 300 km/s) is too low to resolve the
lines. However, some lines appear to be marginally resolved. It is 
possible that weak lines look  broad due to noise spikes in the adjacent continuum, as we think is the case of \Pag\ in object \#3, which has an unrealistic FWHM of 950 km/s; we did not try to correct
for this effect the equivalent width in Table \ref{table_1}, which is hence
very likely overestimated by a factor of $\sim$2. 
In two stars (\#19 and \#21), the He~I line has an inverse P-Cygni profile, with redshifted absorption; the line is strong in \#19, but 
barely detected in \#21.
Table \ref{table_1}  gives the total equivalent width of the line, while the values of the absorption and emission components are given in the notes to the table.

We have been surprised by the number of objects where \Pag\ is detected while
\Pab\ is not.
The ratio of the \Pag\ to \Pab\ fluxes has been recently measured for a group of CTTS in Taurus by 
Bary and Matt (\cite{Bea07}), who find a tight correlation with a ratio
0.86$\pm 0.10$. 
%{\bf 
Magnetospheric accretion models  (Muzerolle et al.~\cite{Mea01})
predict ratios $\sim 0.7-0.9$ for \Macc $\simgreat 10^{-8}$ \Msun/yr,
increasing to $\sim 1.2-1.4$ at \Macc $\sim 10^{-9}$ \Msun/yr
(J. Muzerolle, private communication).
%}
Considering that for TTS the ratio of the continuum
at the two line wavelengths is $\sim 1.2-1.4$, we expect in any case comparable
equivalent widths,
and, in fact, this is the case in the
five \sori\ objects where both lines are detected.  
All stars with \Pag\ but no \Pab\ have relatively weak \Pag; however, in several cases we should have detected \Pab\ if the equivalent widths were
similar and the lines unresolved. 
We have checked the observational and data reduction procedures, and found no 
obvious explanation for the absence of \Pab. It is possible that the upper
limits to \Pab, derived assuming that the line is not resolved, are
in fact underestimated, and, indeed, some objects with very low  limits
to the ratio of the \Pab/\Pag\ equivalent width (e.g., \#3, \#22, \#26)
have broad \Pag\  (see
Fig.~\ref{fig_pag}). Also, in some cases \Pab\ may have a P-Cygni profile 
with deeper absorption than
\Pag, and this  could reduce the ratio of the unresolved equivalent widths.
This kind of profile, however, is quite rare in Taurus TTS (e.g., Edwards et al.~\cite{Eea06}, Folha \& Emerson \cite{FE01}).
Note that the two lines \Pag\ and \Pab\ are observed simultaneously, so that
variability cannot affect their ratio. 
%However, as we will discuss in the following, none of our conclusions depends
%on it. 

\begin{figure}
   \centering
   \resizebox{\hsize}{!}{\includegraphics[]{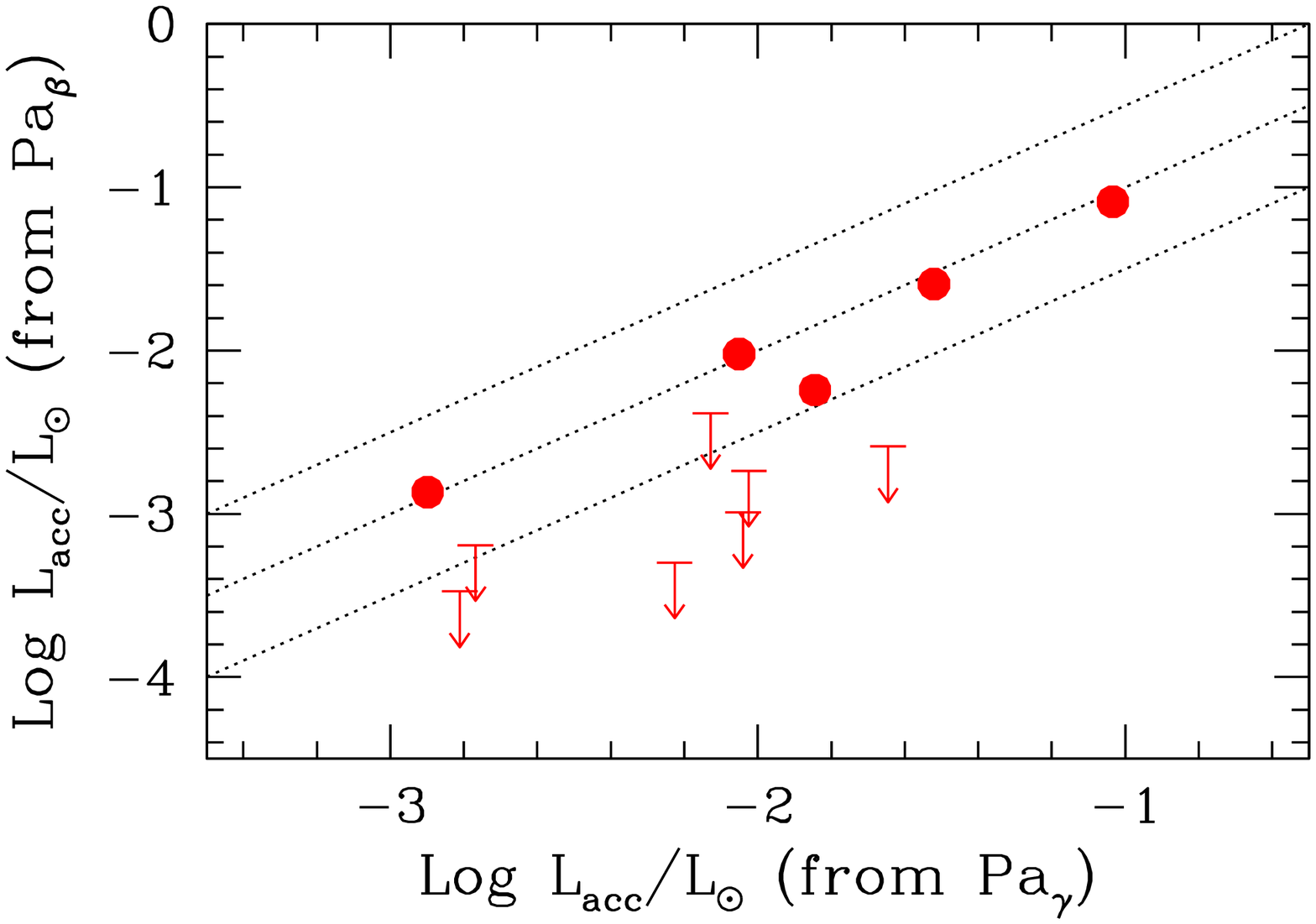}}
	\caption{Comparison of the accretion luminosity computed from \Pag\ 
and that computed from \Pab\ for all objects with \Pag\ detection.
Dots are actual measurements, arrow 3$\sigma$
upper limits. The dotted lines show the locus of \Lacc (\Pag)=\Lacc(Pab) and 
the $\pm 0.5$ range. 
}
\label {fig_lacc_comp}
\end{figure}

\begin{figure}
   \centering
   \resizebox{\hsize}{!}{\includegraphics[]{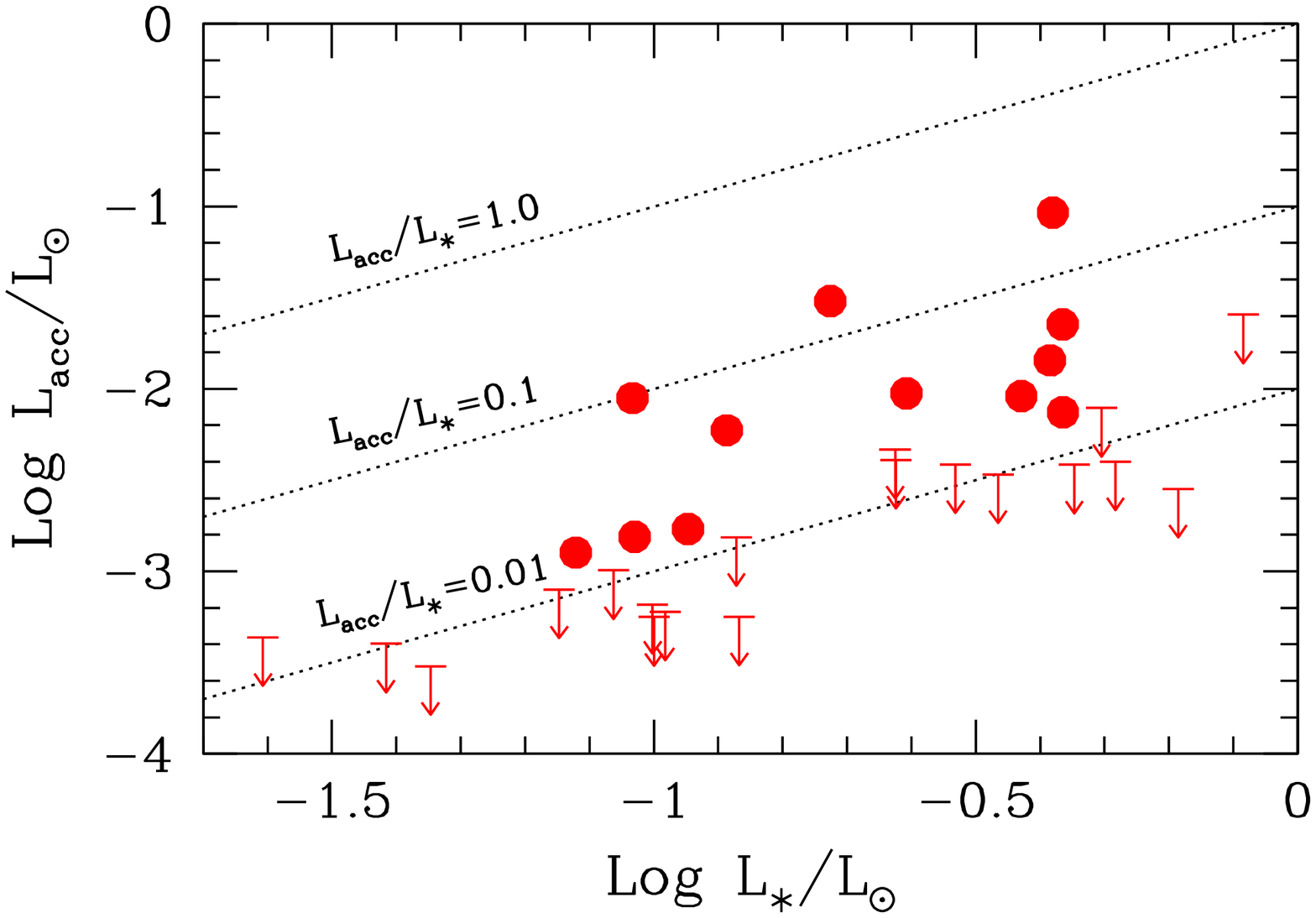}}
	\caption{Accretion luminosity as function of the stellar luminosity
for the Class II objects. Dots are actual measurements, arrow 3$\sigma$
upper limits. The dotted lines show the location of the \Lacc/\Lstar=1.0, 0.1
and 0.01 ratios, respectively.}
\label {fig_lacc}
\end{figure}

The HeI 1.083$\mu$m line is seen in 23/31 Class II stars, in absorption (8 cases), in emission (14 cases), or with a P-Cygni profile (2 cases).
This variety of profiles is known to occurr in TTS
(Edwards et al.~\cite{Eea06}), and  will be discussed  in \S ~4.2.

\section {Discussion}

\subsection {Accretion luminosities and mass accretion rates}

We use the hydrogen recombination lines to  measure the accretion luminosity \Lacc, following the
procedure described in detail by Natta et al.~(\cite{Nea06}).

Firstly, we compute line fluxes from the equivalent widths and the flux of the 
continuum near each line, calibrated with the 2MASS value of the J magnitude
and A$_J$=0.  We apply a correction of a factor 1.3 to the continuum at 1.09 $\mu$m, to account for the typical slope of the J-band continuum in late K -- M stars.

% We have estimated the continuum flux using the 2MASS values of the J magnitudes (see column 7 of Tab.\ref{table_1}), assuming extinction A$_V$=0 (ref), integrating the spectra over the 2MASS J filter profile (obtaining $<N_{2MASS}>$) and scaling to the measured 2MASS value. So we have obtained the fluxes for Pa$\beta$ and Pa$\gamma$ line by means of the following relations: 
%
%
%
%\be \label{eq1} F(Pa\beta)=F_{0,J}\times 10^{\frac{-J}{2.5}}\frac{N}{<N_{2MASS}>}\times \frac{c}{\lambda^{2}_{Pa\beta}}\times Ew(Pa\beta) \ee
%
%and 
%
%\be \label{eq2} F(Pa\gamma)=F_{0,J}\times 10^{\frac{-J}{2.5}}\frac{N}{<N_{2MASS}>}\times \frac{c}{\lambda^{2}_{Pa\gamma}}\times Ew(Pa\gamma) \ee
%

%In the same way we have computed the line fluxes of HeI ($\lambda$=10830$\AA$): results for Pa$\beta$, Pa$\gamma$ and HeI luminosities are shown respectively in column 11, 12 and 13 of Table \ref{tab2}. In the same table last column we have shown the mass accretion rate, $\dot{M}_{acc}$, computed from the luminosity of Pa$\gamma$, L(Pa$\gamma$), using the relation between L(Pa$\gamma$) and the accretion luminosity L$_{acc}$:

Line luminosities are computed from the line fluxes for the adopted distance
D=350 pc. 
We use them to derive a measurement of the accretion luminosity \Lacc,
using the empirical correlation between the luminosity of the near-IR hydrogen 
recoombination lines and \Lacc\ first noticed by Muzerolle et al.(\cite{Mea98})
and used to study the accretion properties of TTS in Ophiuchus by Natta et al.~(\cite{Nea06}). 
The correlation has been established quantitatively for the two lines \Pab\ and \Brg\ in a sample
of TTS in Taurus for which reliable measurements of the accretion luminosity from veiling were available
(Muzerolle et al.~\cite{Mea98}; Calvet et al.~\cite{Cea04}), and extended in the case of \Pab\ to the brown dwarf regime by Natta et al.~(\cite{Nea04}).
We adopt here a relation between \Lacc\ and  \Pag\ luminosity obtained
from the Natta et al.~(\cite{Nea04}) correlation between \Lacc\ and the
\Pab\ luminosity and the 
ratio of \Pag\ to \Pab\ fluxes  of $0.86\pm 0.1$
(Bary and Matt \cite{Bea07}):

\be \label{eq3} Log \frac{L_{acc}}{L_{\odot}}=1.36 \times Log \frac{L(Pa\gamma)}{L_{\odot}} +4.1 \ee

%{\bf 
Eq.(1)  does not depend on any  
assumption on the line origin, but only on the  known empirical 
relation between \Lacc\ and the \Pab\  luminosity  and the
adopted value of the \Pag/\Pab\ ratio.
As discussed at the end of \S 3, in several of our \sori\ objects the observed
ratio \Pag/\Pab\ is higher than the Bary and Matt one and the values
(or upper limit) of \Lacc\ computed from L(\Pab) are correspondingly lower
(see Fig.~\ref{fig_lacc_comp}). We decided to use in the following
\Lacc\ derived from \Pag\ to maximize 
the number of actual measurements; since \Lacc (\Pab) $\simless$ \Lacc (\Pag),
deriving \Lacc\ from L(\Pab) would just  strengthen our conclusions.
%}
%Values of \Lacc\ can be computed independently from the \Pab\ luminosity.
%
%For the 5 objects where both lines are detected, and for 3 additional ones with
%\Pab\ upper limits, the accretion luminosities agree within a factor of 3, i.e.,
%within our estimated uncertainties.

%\begin{figure}
%   \centering
%   \resizebox{\hsize}{!}{\includegraphics[]{fig_ewratio.eps}}
%	\caption{Ratio of \Pab\/\Pag\ equivalent widht as function of the
%\Pag\ equivalent width. Only objects with \Pag\ detections are included.}
%\label {fig_ewratio}
%\end{figure}

%a version suited for Pa$\gamma$ line of the relation $Log \frac{L_{acc}}{L_{\odot}}=1.36 \times Log \frac{L(Pa\beta)}{L_{\odot}} +4$, taken from Muzerolle et al. 1998b and Natta et al. 2002. So we have used L$_{acc}$ computed with Pa$\gamma$, because we have got a larger (12) number of detections respect to Pa$\beta$ line detecitons (5, see also Tab.\ref{tab2}).\\

The mass accretion rate $\dot{M}_{acc}$ is then been computed from $L_{acc}$:
\be \label{eq4} \dot{M}_{acc}=  \frac{ L_{acc}R_*}{GM_*} \ee
where \Mstar\ and \Rstar\ are the stellar mass and radius, respectively.

The uncertainties on \Lacc\ and \Macc\ are difficult to estimate, but they are
certainly not small. In addition to the measurement errors on the line
equivalent widths, there are continuum calibration uncertainties, since our
spectra have not been  photometrically calibrated. Variability in the lines
is a well known phenomenon in TTS, and  snap-shot estimates of their
strength can differ from the long-term average values by large factors.
Moreover,
the relation between L(\Pag) and \Lacc\ has an intrinsic, 
non-negligible scatter,  and the determination of the stellar mass/radius depends on a
number of assumptions, among them the adopted evolutionary tracks.
The final uncertainties on $\dot{M}_{acc}$ can easily be of $\pm 0.5$ dex,
as discussed, e.g., by Calvet et al.~(\cite{Cea04})
and Natta et al.~(\cite{Nea06}).
In spite of all that, snap-shot accretion values of a large number of stars in
any given star formation region provide very significant information on the
typical accretion properties of the region and on its evolutionary stage.

The results for \sori\ are summarized in Fig.~\ref{fig_lacc} and \ref{fig_macc}, top panel.
Fig.~\ref{fig_lacc} plots the accretion luminosity as a function of the stellar
luminosity for all Class II objects. It shows clearly that very few stars
have \Lacc\ larger than 0.1~\Lstar, and that most of them have values well below
this limit. This is different from the results obtained in younger star
forming regions, such as Taurus and Ophiuchus (see summary in Natta et al.~\cite{Nea06}), where a large fraction of stars has \Lacc/\Lstar $>0.1$.

Fig.~\ref{fig_macc} shows the mass accretion rate 
as function of the stellar mass for the same 31 objects. 
The values of \Macc\ are all smaller than  $\sim 10^{-8}$ \Msun/yr, and
23/31 stars have accretion rates lower than $10^{-9}$ \Msun/yr. 
When compared to stars with similar mass in Taurus and Ophiuchus, 
these are very low values, as we will discuss in \S~4.5.

\subsection {He line}

Edwards et al.~(\cite{Eea06}) have recently published a  study
of the \Pag\ and He~I profiles in a sample of 39 Taurus TTS; they found that
the strength of both lines correlate with the J-band veiling, i.e., with the
accretion luminosity.  Although with a large spread, objects with low veiling have on average not only weaker but also
narrower \Pag.
He~I lines have very often P-Cygni profiles, with blue-shifted absorption 
caused by outflowing gas and red-shifted absorption due to infalling material.
On average, strong accretors have the He~I line in emission; red-shifted and blue-shifted absorption features are present in high and low accretors, but,
if unresolved, the He~I equivalent width would turn from strong emission in
high accretors to net absorption in low accretors.
In these objects, the blue-shifted absorption features due to outflowing matter, very likely a non-collimated wind, dominate the profile.

Fig.~\ref{fig_he} plots the equivalent width of the He~I line as function of the
accretion luminosity \Lacc\ for our sample. 
If we consider objects with measured \Lacc, 
we find a good correlation between the two quantities, with He~I in net absorption in low accretors and in emission in high accretors, as in the Edwards et al.~(\cite{Eea06}) sample. 
Note that Edwards et al.~ use in their analysis
the ``activity index" of the He~I line,
i.e., the sum of the absolute values of the
emission and absorption equivalent widths, 
while our unresolved profiles give the net equivalent width only. 
The trend of decreasing emission, increasing net absorption
for decreasing \Lacc\ 
%{\bf 
is weaker
%} 
if  we include objects with \Lacc\ upper limits;
as in Taurus, there are some objects with strong He~I emission
but low \Lacc (e.g., \#27), which it would be interesting to investigate further.

The Taurus sample includes CTTS with mass accretion rates \Macc\ between
$\sim 5\times 10^{-6}$ and $5\times 10^{-10}$ \Msun/yr, with a large fraction
(29/38) of objects accreting at rates higher than the highest values of the \sori\ sample.
Edwards et al.~(\cite{Eea06}) divide their sample of CTTS into four
groups according to the 1$\mu$m veiling. The median values of Log~\Macc\ of the first and second group (high and medium veiling) are -5.8 and -7 \Msun/yr, while the third and fourth (low 1$\mu$m veiling, narrow and broad
\Pag, respectively) have median -8.2 \Msun/yr. The low veiling, broad \Pag\ group
has, on average, He~I lines with net negative equivalent widths. The \sori\ sample
overlaps with the low veiling objects and extends it to even lower values of
\Macc. In fact, there are no objects with
the very strong He~I emission lines of  some Group I Taurus stars, while
the fraction of objects with net He~I absorption is higher than
in the  Edwards et al.~(\cite{Eea06}) group IV. 

An interesting aspect of these results is that absorption He~I lines can be detected even in objects with very low \Macc, where direct accretion signatures 
(e.g., hydrogen line emission, veiling) can be hard to measure.  Accretion-powered winds may be the longest lasting tracers of disk activity.

\subsection {Evolved Disks}

The \sori\ cluster contains a relatively high fraction of evolved disks, i.e., objects 
with IR excess emission lower than the median of CTTS. Some of them
are classified by Hern\'andez et al.~(\cite{Hea07}) as transitional disks,
i.e., objects with a very low excess in the near-IR but a normal excess at longer
wavelengths.
Our sample includes two such objects. 
One (\#28, S897) is classified as a transitional disk; it has 
no \Pab\ or \Pag\ lines above the detection limit (the corresponding mass
accretion rate is $<10^{-9}$ \Msun/yr), but a tentative detection 
of He~I, and a rather broad \Ha\ (10\%FW=504 km/s; Sacco et al.~\cite{Sea07}). 
The other object (\#29, S908) is classified as an evolved disk, 
but shows both \Pab\ and \Pag\ in emission and the He~I line in absorption.
Also for this object, Sacco et al.~(\cite{Sea07}) measure a broad \Ha, with
a 10\%FW=401 km/s.
Its accretion rate is similar to that of the Class II \sori\ stars.

It is possible that in these objects  (\#29 in particular)
the low IR excess is due to the disk orientation on the
plane of the sky. However, some transitional disks are known to have 
relatively large \Macc\ (see summary in Chiang and Murray-Clay (\cite{CM07})
and references therein), and the discussion on how long accretion can be
sustained once disks start to evolve is still open.
The relatively large sample of evolved and transitional disks in \sori\ is 
well  suited to study the accretion properties of transition and evolved
disks. However, the sample discussed in this paper
includes too few such objects to provide any answer.

\subsection {Accretion Evolution}

%In addition, the 2MASS J magnitudes are not measured simultaneously to our 
%spectra, so that line fluxes are not just a 
%time snapshot of the stellar activity, but also the product of two quantities 
%that may vary independently. 

%\be \label{eq4} dot{M}_{acc}=L_{acc} \times \frac{R_*}{GM_*} \ee

\begin{figure}
   \centering
   \resizebox{\hsize}{!}{\includegraphics[]{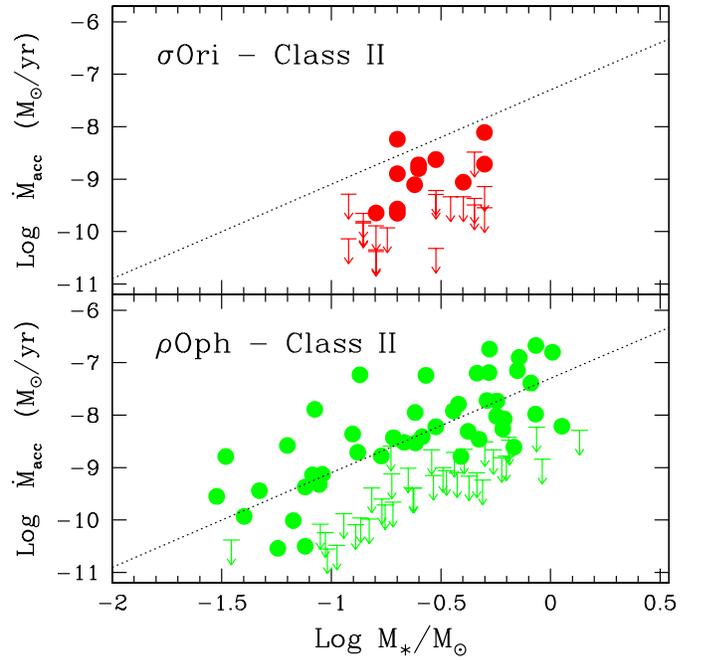}}
	\caption{Mass accretion rate as function of \Mstar\
for \sori\ (top) and \roph\ (bottom; from Natta et al.~\cite{Nea06}).
Dots are actual measurements, arrows 3$\sigma$ upper limits. The dotted line
(same in both panels)  shows the  $\log{M_{acc}} = -7.3 + 1.8\times \log{M_\star}$ locus,
that we take as a fiducial line to define "high accretors" (see text).
}
\label {fig_macc}
\end{figure}

\begin{figure}
   \centering
   \resizebox{\hsize}{!}{\includegraphics[]{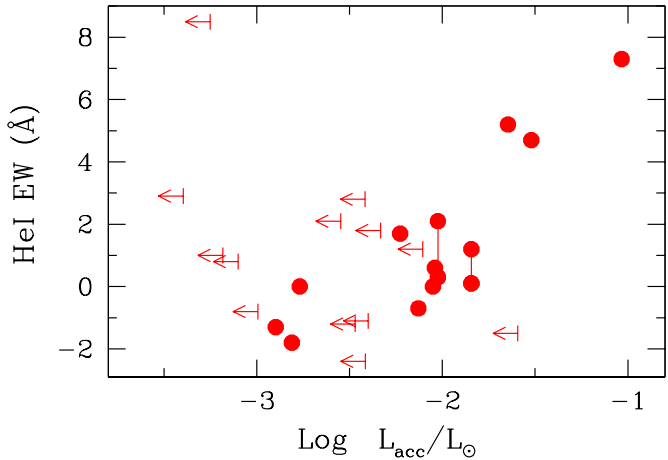}}
	\caption{HeI equivalent width as function of \Lacc\ for Class
II objects. Dots are objects with an actual measurements of \Lacc, while arrows
are upper limits. For clarity, we do not plot the objects with \Lacc\ upper limits and non-detection of the He~I line.
For the two cases where we resolve the P-Cygni profile
of the HeI line, we show (connected by a line) the corresponding unresolved
equivalent width  and the HeI activity index (sum of absolute values of absorption and emission equivalent widths; Edwards et al.~\cite{Eea06}).}
\label {fig_he}
\end{figure}

\begin{figure}
   \centering
   \resizebox{\hsize}{!}{\includegraphics[]{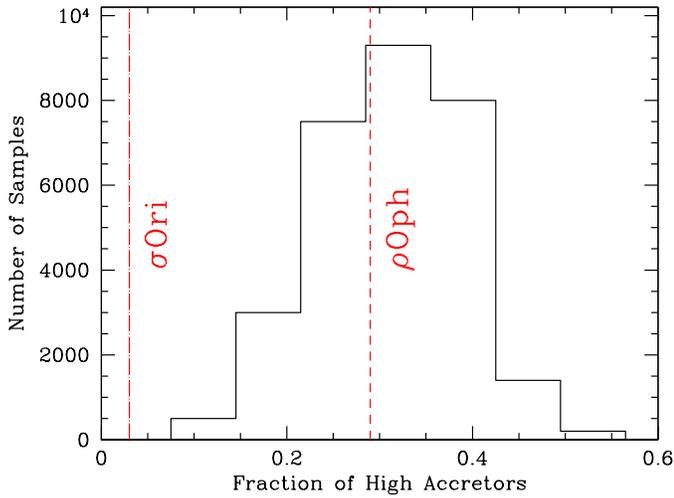}}
	\caption{Distribution of simulated fraction of high accretors in samples extracted from a parent distribution ias measured in \roph.
%Number of runs as function of the fraction of high accretors
%(see text for the definition) in each run. 
The dashed line shows the
fraction of high accretors in \roph, the dot-dashed line that in \sori.  }
\label {fig_2pop}
\end{figure}

\begin{figure}
   \centering
   \resizebox{\hsize}{!}{\includegraphics[]{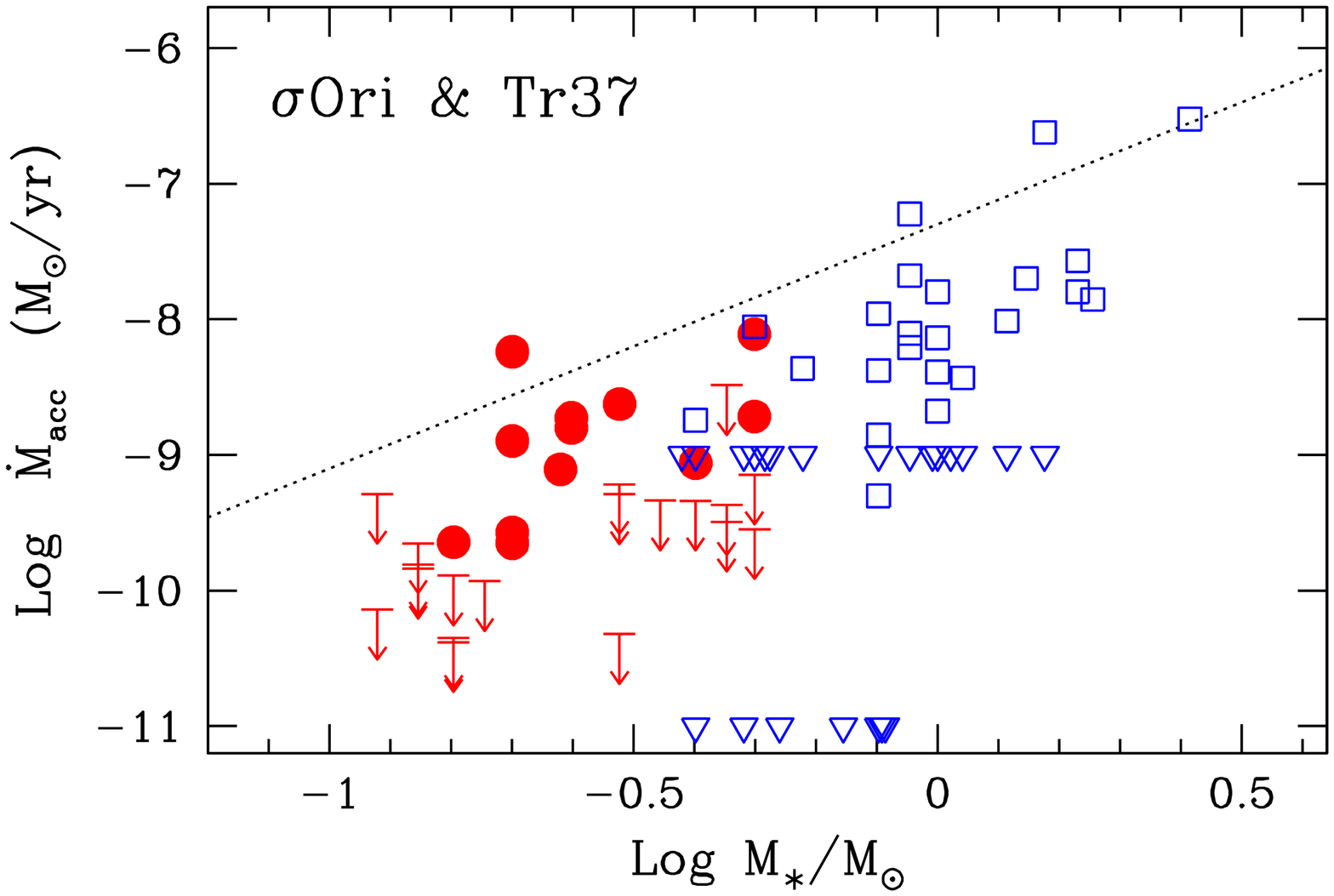}}
	\caption{\Macc\ vs. \Mstar\ for \sori (dots and arrows; same as Fig.~\ref{fig_macc}) and Tr37 (from Sicilia-Aguilar et al.~\cite{SAea06b}).
Squares shows value of \Macc\ derived from U-band photometry, triangles at
$\log$\Macc=-9 \Msun/yr are objects with no U-band excess but broad \Ha, triangles at 
$\log$\Macc=-11 are objects with no U-band excess and narrow \Ha. Note that
for this group Sicilia-Aguilar et al.~ estimate $\log$\Macc$<-12$, we plot them at -11 for convenience. }
\label {fig_mold}
\end{figure}

%As indicated in \S~4.1, \sori\ TTS have accretion rates lower on average than
%their counterparts in younger star forming regions, such as Taurus and
%\roph.  
The relevance of the \sori\ results for the understanding of the accretion
evolution can be better understood in comparison with  other regions of star formation, both younger and older.

The comparison with  the \roph\  star forming region is particularly
interesting. Mass accretion rates in \roph\ have been derived by Natta
et al.~(\cite{Nea06}) for an IR-selected sample of Class II complete
to a limiting mass of about 0.03 \Msun, using the luminosity of near-IR
hydrogen recombination lines as a proxy of the accretion luminosity.
Because the  selection criteria  and methods are similar, the comparison
between these two regions is more meaningful than in other cases.

The major difference between the two regions is the lack in \sori\
of objects with high accretion rates. If we restrict the comparison 
to the mass range 0.1--0.5 \Msun, where the two samples overlap, 
and take as a reference the line
shown in Fig.~\ref{fig_macc}, which is the best fit to the \roph\
sample of measured \Macc\ over the whole mass interval, we find that
only 1 star in \sori\ lies above this line (3\%), while the fraction is
$\sim$30\% in \roph.

%A more quantitative check is shown in Fig.~\ref{fig_2pop}, which shows the
%probability that the two samples are extracted from the same parent population.
%The probability is very low, and can be ruled out.{\bf details: check that
%the fiducial line is the same as in the figure; did LT select J<14.5 mag
%objects? did LT consider only the restricted range of masses?}

% 
% Text for MC simulations
%

Because the \sori\ sample is small (much smaller than the \roph\ one), 
we performed a
simplified Monte Carlo simulation 
to assess the statistical significance of this difference.
The goal of our simulation is to evaluate 
whether the number of high accretors we observe in the \sori\ sample is 
consistent with being drawn from  a distribution of rates as derived from the
\roph\ Class~II objects observations. 
We take the \roph\ stars in the interval
0.1--1 \Msun\ 
%{\bf 
(the sample  0.1--0.5 \Msun\ would have too few sources
to derive a reliable \Macc\ distribution)
%} 
and use the relationship between accretion rates and
stellar mass derived by Natta et al.~(\cite{Nea06}) to scale all measurements to
a fiducial stellar mass of 1~M$_\odot$;
we then compute the fraction of high
accretors as the ratio of objects with measurements above the fiducial line
shown in Fig.~\ref{fig_macc}
to the total (measurements plus upper limits). Within the chosen
mass range, the fractions of high accretors in \roph\ and \sori\ 
are 29\%\ and 3\%\ respectively. 

Using a Monte Carlo method, we extract a large number of samples of 31
accretion rates. For each sample we randomly remove a number of objects 
based on the 
fraction of upper limits in  \roph\  ($\sim$52\%), while
for the remaining objects we simulate measurements using
the distribution of measured rates.
Figure~\ref{fig_2pop} summarizes the results:
% of our simulations of the number of high accretors: 
the histogram plots the distribution of the fraction of high accretors
in the simulations, the dashed line shows the value for the \roph\
sample while the dashed-dotted line shows  the value observed in our \sori\ sample.
The value observed in \sori\ is lower than the 0.1 percentile of the 
Monte Carlo simulations, suggesting that it is  very unlikely 
that the two samples are two different realizations of the same population.

The lack of strong accretors in \sori\ is very likely an age effect, as
in viscous disks
the mass accretion rate is expected to decrease with time (roughly as
$t^{-1.5}$; Hartmann et al.~\cite{Hea98}).
On average, this means approximately a factor $10$ 
from the young (less than 1 Myr) \roph\ population to the older ($\sim$ 3 Myr)
\sori\ one, consistent with the results shown in Fig.~\ref{fig_macc}.
One way of looking at this is to compute the fraction of high accretors
in \sori\ not with respect to the ``younger" \roph\ \Macc-\Mstar\ fiducial
line, but
with respect to an ``aged" version, where
\Macc\ has decreased by the
same factor 10 for each \Mstar. Then,  the fraction of \sori\ high accretors
will be  of $\sim$30\%, identical to the fraction of
high accretors in \roph.

Fig.~\ref{fig_mold} shows a comparison between mass accretion rates
in \sori\ and in the $\sim$4 Myr old star forming region Tr~37.  This region, located at a distance of $\sim 900$ pc, has been extensively studied by
Sicilia-Aguilar et al.~(\cite{SAea04}, \cite{SAea05}, \cite{SAea06a}, \cite{SAea06b}), who have 
classified disk properties from {\it Spitzer} IRAC and MIPS data and measured
\Macc\ from U band excess and \Ha\ profiles. 
The methods of deriving
\Macc\ are different in the two regions, but probably any systematic
effect is within the uncertainties, since both  correlations between
the U-band excess and \Lacc\ and  between the hydrogen IR line luminosity
and \Lacc\ have been empirically calibrated using the same sample of TTS in 
Taurus with good measurements of \Lacc\ from veiling 
(e.g., Calvet and Gullbring \cite{CG98}; Muzerolle et al.~\cite{Mea98}). 

Fig.~\ref{fig_mold} plots all the
stars with disks (bone-fide Class II and transition disks) with measured stellar mass and \Macc, as reported by Siciliar-Aguilar et al.~(\cite{SAea06b}). 
We have not included  any of the
very young ($\sim 1$ Myr)
Class II objects, probably an episode of triggered star formation, found
in the TR~37 ``globule".
Because of its larger distance,
the Tr~37 sample  covers a mass range between $\sim 0.4$ and 1.8 \Msun, with 
one star of 2.5 \Msun (whose \Macc\ is very uncertain); 
there is practically no overlap with the \sori\ lower-mass sample. 
However,
the accretion properties of the two regions look very similar.
The trend of higher \Macc\ for higher \Mstar\ is confirmed, with  good merging between the two samples. More importantly, both \sori\ and
Tr~37 lack high accretors to a similar degree. If we define, as before, 
high accretors to be those stars with values of \Macc\ above the fiducial line s
hown in Fig.~\ref{fig_macc}, the fraction of high accretors in Tr~37 is 6\%,
similar, within the uncartinties, to the 3\% found in \sori, 
and certainly much lower than the 30\% fraction of high accretors in \roph. 
%It should be noticed that the
%Tr~37 sample includes a small fraction of much younger stars (age $\sim 1$ Myr; the so-called globule), which are probably an episode of triggered star formation. Only 5 of them are plotted in Fig.~\ref{fig_mold}, as
%there are no information on \Macc\ and/or \Mstar\ for the others; only
%one of them is above the fiducial line, so that excluding them would not change the results. 

The continuity in the \Macc--\Mstar\ behaviour between the two regions
seems to extends to stars more massive than
about 1--1.2 \Msun. In other star-forming regions, as discussed in \S 1,
stars more massive than $\sim 1$ \Msun\ 
loose evidence of disks and accretion much faster than less massive stars, and one could expect to see a drop 
in the fraction of Class II stars 
among the more massive members of Tr~37.
However, this does not seem to be the case,
because Tr~37 has a relatively high fraction of Class II (8 with respect to 5 Class III among stars more massive than 1.2 \Msun), even when the globule
population is not taken into account. Of these, 6 have evidence of accretion.
%It would be of great interest to study the accretion properties across
%the mass spectrum of other, relatively old regions, such as, e.g.,
%U Sco or NGC~2362, where  the fraction of disks decreases
%significantly among massive stars (Carpenter et al.~\cite{Cea07};
%Dahm and Hillenbrand \cite{DH07}).
The case of Tr~37 shows that the evolution of disks and accretion
as function of the stellar mass is still an open problem, that needs to be addressed in more detail in the future.

%The results of these two relatively old regions confirm that
%the accretion rate decreases with time, roughly as expected
%from viscous disk models, over a rather large range of masses.
%This decrease occurs steadily as disks
%age, for a time comparable with the lifetime of the dusty disks themselves.

%\subsection {The effect of the bright \sori\ star}

\section {Summary and conclusions}
This paper presents J-band spectra of 35 stars in the star-forming region
\sori. We concentrate our attention on the three lines that are indicative of
accretion-related activity, namely the two hydrogen recombination line
\Pab\ and \Pag\ and the He~I line at 1.083 $\mu$m.

The stars range in mass from $\sim 0.1$ to $\sim 0.5$ \Msun.
Four have no evidence of disks (Class III), according
to the classification of Hern\'andez et al.~(\cite{Hea07}) based on
{\it Spitzer} near and mid-IR photometry. None of the three lines is detected
in their spectra, with the possible exception of one star which has a 
tentative detection of He~I emission. The other 31 objects have associated disks (Class II and
evolved disks), and we detect \Pag\ in 12 of them and \Pab\ in 5.
The He~I line is seen in 23 disk objects, either in emission or in absorption or both.
Even at the low spectral resolution of our spectra ($\sim 300$ km/s), in 2 cases
the He~I has a P-Cygni profile with redshifted absorption.

We derive accretion luminosities and mass accretion rates from the \Pag\
luminosity;
%, as discussed in Muzerolle et al.~(\cite{Mea98}), Calvet et al.~(\cite{Cea04}), Natta et al.~(\cite{Nea04}) and used in the large survey of
%\roph\ pre-main sequence objects by Natta et al.~(\cite{Nea06}).
the results for \sori\ indicate values much lower than in stars of similar
mass in the younger regions Taurus and \roph. In particular, a statistical
analysis shows that the \roph\ and \sori\ populations cannot be drawn
from the same parent population. TTS in \sori\  are statistically
identical to those
in \roph\ only if we take into account their age, i.e., that
any \roph\ object will have, at \sori\ age,  a value of \Macc\ lower by 
a factor of about 10 (\Macc$\propto t^{-1.5}$), as expected if viscosity  controls disk evolution.

The HeI is detected (in net absorption or emission) in more stars than either
of the hydrogen lines. Its behaviour agrees qualitatively
with what found by
Edwards et al.~(\cite{Eea06}) in a sample of TTS in Taurus;
namely, strong HeI emission characterizes on average high accretors, while
net absorption, probably due to a wind, is typical of low accretors.
We extend this result to objects of very low \Macc\ and suggest that the He~I
line may be used to detect weak accretors when other methods become unfeasible.

The \sori\ region, with its well studied disk population and wealth of data at all available wavelengths, 
large number of member stars extending over a broad
mass range and moderate distance, is particularly well suited for a number 
of follow-up studies. Among them, we want to mention that it will be relatively 
easy to obtain near-IR spectra of the complete sample of Class II objects, 
down to the brown dwarf regime, to check, e.g., if the accretion rate 
evolves differently in brown dwarf disks.
The comparison of \sori\ with the region Tr~37, which has similar age
and a comparable fraction of high accretors, 
but a surprisingly high fraction of disks among stars more massive than about 1 \Msun,  
shows that the dependence of disk and accretion evolution with the mass
of the central object is still an open and intriguing problem.

Our results on the two evolved/transition disks in our sample
confirm that accretion is still going on in some of them
and suggest that near-IR spectroscopy and high resolution spectra of the
hydrogen and He~I lines may help to clarify the last stages of the
history of accretion disks.

\begin{acknowledgements}
We thank James Muzerolle for providing us with unpublished results of his magnetospheric accretion models and for very useful discussions.
It is a pleasure to acknowledge the continuous, competent
and friendly support of the ESO staff during the preparation
and execution of the  observations at
La Silla observatory.
This project was partially supported by MIUR grant
2004025227/2004 and by the INAF 2006 grant ``From Disks to Planetary Systems".
\end{acknowledgements}

{}

\end{document}